\documentstyle[prb,twocolumn,aps,epsf,epsfig]{revtex}

\newcommand{\rvec}{{\bf r}}

\newcommand{\xvec}{{\bf x}}
\newcommand{\beq}{\begin{equation}}
\newcommand{\eeq}{\end{equation}}
\newcommand{\beqa}{\begin{eqnarray}}
\newcommand{\eeqa}{\end{eqnarray}} 

\begin{document}
\twocolumn[\hsize\textwidth\columnwidth\hsize\csname
@twocolumnfalse\endcsname
\draft
\title{The induced charge in a Fr\"ohlich polaron: Sum rule and 
spatial extent.} 

\author{Sergio Ciuchi$^1$ Jose Lorenzana$^{2,}$\cite{adr} and Carlo Pierleoni$^{1,2}$}

\address{{$^1$}
INFM and Dipartimento 
di Fisica, Universit\`a de L'Aquila,\\
via Vetoio, I-67100 Coppito-L'Aquila, Italy.}

\address{{$^2$}
INFM, UDR Roma I,
Universit\`a di Roma "La Sapienza",\\
Piazzale A. Moro 2, I-00185 Roma, Italy.}

\date{\today}
\maketitle

\begin{abstract}
Within the path integral formalism, we derive exact expressions for 
correlation functions  measuring the lattice charge induced by an 
electron and associated polarization in  Fr\"ohlich polaron problem.
We prove that a sum rule for the total induced charge, already obtained within
approximated approaches is indeed exact.
As a consequence the total induced charge is shown rigorously to be
temperature independent.
In addition we perform  path integral Monte Carlo calculations
of the correlation functions and we compare with variational results based on
 Feynman  method.  
As the temperature increases the polaron radius decreases. On the other 
hand at high temperatures the electron motion is not hindered by the lattice. 
These apparently contradictory results are discussed. 
\end{abstract}
\vskip2pc]

\narrowtext
\section{Introduction}
An electron added to an insulating polar crystal forms a 
quasiparticle called dielectric polaron after Fr\"ohlich\cite{kup62,dev96}. 
This  has been recognized as a fundamental field theoretical 
problem\cite{fey65}.
More recently a variety of novel materials have emerged which 
present interesting properties when doped away from an insulating 
phase, like colossal magnetoresistent manganites and high temperature
superconducting cuprates. The fact that these are polar crystals has 
produced a renewed interest in Fr\"ohlich polaron
problem\cite{emi89,iad96,fra98,lor99}.

Roughly speaking
a dielectric polaron is composed of an electron and the (opposite) charge 
that it induces in the lattice. Electron and induced charge attract
each other
so that for the electron to move it has to drag the induced charge resulting
in an increase of the quasiparticle mass. In this work we 
study correlation functions which measure the magnitude and  spatial extent of 
the induced charge and associated polarization field.

We derive a rigorous  sum rule which   states that  
the total induced charge equals the charge induced by a classical (static) 
electron and it is independent of temperature.
This result, well known within perturbative \cite{pin62}
and variational approaches \cite{pee83}, is proven here to be exact.
The large distance behavior of the electric field
is determined by this sum rule. 
In addition we discuss the short distance and the high temperature 
asymptotic limit of these quantities.  
These   results  provide constrains to  approximations on the polaron
problem. 

Real space  path integral  Monte Carlo (PIMC)
method is used to evaluate correlation functions. 
These are compared with Feynman's variational
approximation (FVA) \cite{pee82,pee83}
and  analytical results in weak and strong 
coupling. We find that the polaron radius is determined at low
temperatures by the electron-phonon coupling $\alpha$ alone while at high 
temperatures it is proportional to the  de Broglie  thermal wave 
length  ($\lambda_T=\hbar \sqrt{2\pi\beta/ m}$ 
with $\beta$ the inverse temperature)
and becomes independent of coupling. We define a polaron 
crossover temperature $T^*(\alpha)$.
Although the electron induces a temperature independent 
charge in the lattice the induced charge hinders the electron motion
only  below $T^*(\alpha)$. At high temperatures thermal effects wash
out the hindering effect of the induced charge but, we remark again,
not the induced charge itself. 

\section{Analytical Results}

Our starting point is the effective action for the Fr\"ohlich polaron problem,
after the phonons have been eliminated with the usual path integral
techniques\cite{fey65}, $S=S_0+S_I$ with
\begin{eqnarray}
  \label{eq:eff-act}
  &S_0[\xvec]& = \int_0^{\hbar \beta} d\tau \frac{1}{2}m \dot{x}^2\\
&S_I[\xvec]&=-\frac{\alpha}{2\sqrt{2}} \frac{(\hbar \omega_L)^{3/2}}{m^{1/2}}
\int_0^{\hbar\beta} d\tau_1 \int_0^{\hbar\beta} d\tau_2 
\frac{ D(\tau_1-\tau_2)}{|\xvec(\tau_1)-\xvec(\tau_2)|}\nonumber
\end{eqnarray}
here $m$ is the electron mass,
$\alpha= e^2 m^{1/2}/\hbar \bar \epsilon (2\hbar \omega_L)^{1/2}$,  
is the coupling constant with $\omega_L$ the phonon frequency,
$ 1/\bar \epsilon=1/\epsilon_{\infty}-1/\epsilon_0$
and  $\epsilon_{\infty}$ ($\epsilon_0$) is the high (zero) frequency
 dielectric constant. 
 \begin{equation}
   \label{eq:ddtau}
 D(\tau)=\frac{\exp(\omega_L |\tau|) + \exp[(\hbar\beta-|\tau|)\omega_L]}{
\exp(\hbar\beta\omega_L)-1} 
 \end{equation}
is the phonon propagator.

We are interested in the correlation function between the electron
charge density  $-e \tilde{n}({\bf r})$ and the charge induced in the lattice
 $e \tilde{n_i}({\bf r})$ normalized to the probability density to find an electron
at a given point i.e. the inverse of the volume $V$. 
Dropping the charges this is defined as,
\begin{equation}
  \label{eq:gdn}
\langle \tilde{n}(0)  \tilde{n_i}({\bf r}) \rangle /  V^{-1}\equiv
 g(r)/\bar \epsilon
\end{equation}
Averages are defined as path integrals weighted by $S$
\begin{equation}
\label{eq:average}
\langle ... \rangle=\frac{\int {\cal D\xvec} e^{-S/\hbar} (...)}
{\int {\cal D\xvec} e^{-S/\hbar}}
\end{equation}
where the paths entering in
Eq.~(\ref{eq:average}) depart and arrive at the same point.
Further integration over such a point is not performed and 
this assigns to Eq.~(\ref{eq:average}) the 
meaning of a constrained average with $\xvec(0)=\xvec(\beta)=0$.
Those averages are however equivalent to the unconstrained 
ones because of translational invariance.
We used the spherical
symmetry of the problem to define $g(r)$, and we divided  by $\bar \epsilon$
 on the right hand side of Eq.~(\ref{eq:gdn}) for later convenience. 
 Another quantity of interest is the induced lattice polarization 
 \begin{equation}
  \label{eq:defp}
  {\bf P}({\bf r}) \equiv \langle \tilde{n}(0)  \tilde{\bf P}({\bf r}) 
\rangle /  V^{-1} 
 \end{equation}
${\bf \nabla} \tilde{\bf P}({\bf r})=-e \tilde{n_i}({\bf r})$ is the density of polarization
operator.  
The correlation function in Eq.~(\ref{eq:defp}) is related to 
the induced electrostatic potential 
[$\nabla^2V({\bf r})=-4\pi eg(r)/\bar\epsilon$] considered in
Ref.~\onlinecite{pee85} and which will  not be discussed here.
We stress that these quantities have the meaning of {\em correlation 
functions} measuring average induced charge, polarization and induced potential
 at a distance $r$ from the electron position. 

The charge density operator for the  phonons is\cite{mah90},
\begin{equation}
 \label{eq:nidnq}
 e \tilde{n_i}({\bf r})=-
 \sqrt{\frac{\hbar \omega_L}{4  \pi V \bar\epsilon } } 
\sum_{\bf k}  k \tilde Q_{\bf k} e^{i {\bf k}.{\bf r} }.
\end{equation}
where $\tilde Q_{\bf k}$ is the dimensionless displacement for momentum ${\bf k}$
 phonons. Inserting Eq.~(\ref{eq:nidnq}) in Eq.~(\ref{eq:gdn}) we obtain 
an equation for $g(r)$ as a function of the density displacement correlation 
function  $\langle \tilde{n}(0) \tilde Q_{\bf k} \rangle $.
The phonon variables can be traced out by standard methods.
We have found that it is possible to give an {\em exact} expression
for the correlation functions in terms of path integrals weighted by 
the effective electronic action of Eq.~(\ref{eq:eff-act}). 
We find for the density-induced-density correlation function
\beq
\label{eq:gdu}
g(r) = 
\int_0^\beta d\tau U(\tau) 
\left<\delta[\rvec-\xvec(\tau)] \right>
\eeq
and for the polarization field 
\beq
\label{eq:Cp1}
{\bf P}({\bf r}) = -\frac{e}{4\pi \bar\epsilon }  
\int_0^\beta d\tau U(\tau) \left<\frac{\hat {\bf r}}{|\rvec-\xvec(\tau)|^2}
\right>
\eeq
where $\hat {\bf r}\equiv{\bf r}/r$ and
\beq
\label{eq:defU}
U(\tau)=\hbar\omega_L \frac{\sinh(\omega_L \tau)+\sinh[\omega_L (\hbar\beta-\tau)]}
{2 \tanh(\beta\hbar\omega_L/2)\sinh(\beta\hbar\omega_L )}.
\eeq
Within FVA, the variational quadratic action can be exploited in 
Eqs.~(\ref{eq:gdu}),(\ref{eq:Cp1}) to analytically perform the averages
and to recover the results of Refs. ~\onlinecite{pee82} 
and ~\onlinecite{pee85}.

Eqs.~(\ref{eq:gdu}),(\ref{eq:Cp1}) have a simple physical interpretation.
The induced charge can be seen as ``distributed'' along the electron path
with weight $U(\tau)$.  The polarization is the 
superposition of polarizations associated with those elementary induced
charges.

 Eq.~(\ref{eq:gdu}) can be integrated in the whole space using the 
properties of the Dirac's $\delta$ function. 
Since  $\int d\tau U(\tau)=1$, 
we  conclude that $g(r)$ is normalized to one.
The total induced charge $q$ is computed by integrating the
density-induced density correlation function in  Eq.~(\ref{eq:gdn}):
\begin{equation}
  \label{eq:qdg}
  q=e  \int_0^\infty dr 4\pi r^2  g(r)/ \bar \epsilon=\frac{e}{\bar\epsilon} .
\end{equation}
which completes the proof of the sum rule. The total induced charge 
amounts to the charge the electron  would induce  if it where a 
static classical particle. In other words there are no quantum corrections
to the total induced charge. 

An alternative derivation can be work out following Qu\'emerais\cite{que95}. 
From the time derivative 
$i\hbar\ddot{\tilde {\bf P}}= [\dot{\tilde {\bf P}},\tilde H]$ one obtains
\begin{equation}
  \label{eq:dotdotp}
  \ddot{\tilde {\bf P}}=\omega_L^2\left(\frac1{4\pi \bar \epsilon} \tilde {\bf D}
- \tilde {\bf P}\right)
\end{equation}
Here $\tilde H$ is the Hamiltonian, $\tilde {\bf D}$ is the electric 
displacement operator due to
the electron ($\nabla.\tilde {\bf D}=-4 \pi e \tilde n$) and 
$ i\hbar\dot{\tilde {\bf P}}=[\tilde {\bf P},\tilde H]$.
Taking the divergence we obtain a relation for the charge operators:
\begin{equation}
  \label{eq:dotdotp1}
  \nabla.\ddot{\tilde {\bf P}}({\bf r})=\omega_L^2 e
\left(-\frac{\tilde{n}({\bf r}   )}{\bar \epsilon}
+  \tilde{n_i}({\bf r})\right)
\end{equation}
We can integrate this expression in all space and take the 
 thermodynamic average. The left hand side is proportional to the average
of the net force felt by the lattice at the boundary of the system which
should vanish at equilibrium.  The right hand side gives Eq.~(\ref{eq:qdg}).

Eq.~(\ref{eq:qdg}) shows   that 
the induced charge is independent of temperature. This contradicts the 
naive argument that all polaron effects should disappear at high temperatures.
To understand this  behavior one can do an analogy with the behavior of 
an harmonic oscillator in an external field. In that case, because of
 harmonicity, one gets a 
displacement which is temperature independent. Here roughly speaking  the 
harmonic oscillator represents the phonon coordinates and the external field 
is the field produced  by the electron on the phonons. The induced charge is
 a measure of how much the ions displace from their bare equilibrium 
positions in the presence of the electron.  As for the 
single harmonic oscillator, this ``displacement'' is  independent of 
temperature. Only anharmonicities can make the induced charge 
temperature dependent.

Using the sum rule it is easy to see that at distances much larger than
the polaron radius, as defined below, the polarization field goes as
${\bf P}({\bf r})= - e \hat {\bf r}/(4\pi \bar \epsilon r^2)$. Clearly the
distortion produced by the electron is long range, a fact that 
is not always recognized in the literature\cite{ale99}. 
The total electric field (always in the sense of a correlation function)
is given by ${\bf E}={\bf D}-4\pi{\bf  P}$
where we should include in 
${\bf D}=-e  \hat{\bf r}/(\epsilon_{\infty} r^2) $ the
high frequency screening. At long distances we have 
${\bf E}({\bf r})= -e  \hat{\bf r}/(\epsilon_0 r^2)$ which means
that the electric field generated by the electron gets screened by the static
dielectric constant. This is generally expected but to the best of
our knowledge has never been proven for all coupling
and temperatures.

Now we discuss the short distance behavior. 
At distances much  smaller than the polaron radius we expect 
 that the effect of the interaction becomes irrelevant in the 
functional integrals. This is because 
the latter are dominated by electron paths with 
short wave length or equivalently  high kinetic energy. 
We can then 
replace the total action by the free electron action in  
Eqs.~(\ref{eq:gdu}),(\ref{eq:Cp1}). We obtain the asymptotic result:
\beq
\lim_{r\rightarrow 0} {\bf P}({\bf r}) =
 -\frac{e}
{8\pi \bar\epsilon l^2\tanh(\beta \hbar\omega_L/2)} 
{\hat {\bf r}}. \label{eq:Cp1res}
\eeq
where $l=\sqrt{\hbar/2m\omega_L}$ is the 
harmonic oscillator characteristic length. 
Using the same argument we obtain that $g(r)\propto r^{-1}$ for 
$r\rightarrow 0$
and the proportionality coefficient can also be obtained with the same method. 
The latter behavior has been obtained in Ref.~\onlinecite{pee83} within the 
FVA.  These results coincide with lowest order perturbation theory.  

At high temperatures we also expect that the effect of the interaction becomes
irrelevant and so we can replace again 
the total action by the free electron action in the functional 
integrals. The high temperature asymptotic result for $g(r)$ is
\beq
\label{eq:ghiT}
4\pi r^2 g(r) =\frac{2r}{l^2 \beta \hbar \omega_L  } \exp
\left(
 \frac{-r^2}{l^2 \beta \hbar \omega_L}   
\right)\;\;\;\;\beta\hbar\omega_L<<1  .
\eeq
This result has also been obtained in Ref.~\onlinecite{pee83} within FVA. 
We remark that although the density-induced density correlation function does 
not vanish for large temperatures the polaron effective mass tends to the bare
 electron mass.

\section{Numerical Results}

Now we discuss the spatial extent of the induced charge at general 
couplings and temperatures. We have evaluated 
averages in Eq.~(\ref{eq:gdu}) using PIMC.
Eqs.~(\ref{eq:gdu}),(\ref{eq:Cp1}) being expressed in real
space rather that in Fourier components are  more suitable for this purpose.
We have performed Metropolis PIMC calculations within the imaginary time
discretization scheme. In order to regularize the attractive divergence of the
retarded action at short distance, and to improve the convergence with the
number of the imaginary time discretization points,  
we have developed a preaveraging procedure  similar to the one used for local 
actions\cite{dol90,cep95}.
Details of the method and more extensive results will be given in a
separate publication\cite{clp00}.
Here we just say that the results for the $g(r)$ are well
converged as checked by doubling the number of imaginary time slices.
Previous MC studies of the Fr\"ohlich polaron were limited to small 
($\alpha\leq 4$)\cite{ger83} or intermediate ($\alpha\leq 7$)\cite{ale92}
 couplings and were focused to the calculation of the ground state 
energy\cite{ger83,ale92} and effective mass\cite{ale92}.

Following Ref.~\onlinecite{pee83}  we have also computed $g(r)$ in the 
FVA, i.e. using  Feynman's quadratic action to 
evaluate the average  appearing in Eq.~(\ref{eq:gdu})\cite{cas83-84}. 

\begin{figure}[htbp]
   \epsfxsize=8cm
   \epsfysize=5cm
$$
\epsfbox{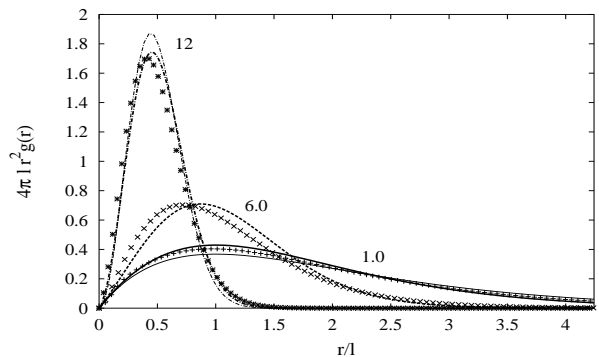}
$$
\caption{
$4\pi r^2 g(r)$ as a function of $r$ for different couplings and 
approximations. From left to right bold curves (FVA
after the formalism of Ref.~\protect\onlinecite{pee83}) and points (PIMC) 
corresponds to $\alpha=12,6,1$, respectively, and $\beta\hbar\omega_L=20$.
Thin solid line is perturbation theory, thin dashed line the Landau-Pekar
strong coupling approximation both at $\beta\hbar\omega_L=\infty$.
}
 \label{fig:gdr}
\end{figure}

In Fig.~\ref{fig:gdr} we show $4\pi r^2 g(r)$ for weak, intermediate 
and strong coupling. We also show the weak coupling result obtained by 
perturbation theory at $T=0$ and the strong coupling result in the Landau-Pekar
approximation. 
For all couplings the correlation function decays exponentially
with distance as expected for a polaron. The area under the curves is
one according to the sum rule Eq.~(\ref{eq:qdg}).
 The short distance asymptotic  behavior Eq.~(\ref{eq:Cp1res}) is exactly 
satisfied in FVA and is also satisfied in the PIMC within the numerical error. 
We define the polaron  radius $r_m$ as the distance at which 
$4\pi r^2 g(r)$ is  maximum.
As the coupling increases the polaron shrinks indicating the 
progressively more localized nature.

\begin{figure}[htbp]
   \epsfxsize=8cm
   \epsfysize=5cm   
$$
\epsfbox{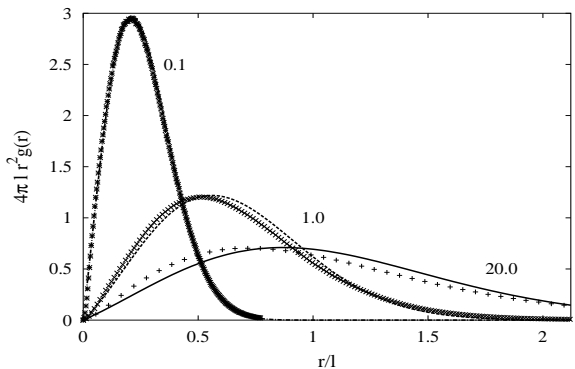}
$$
\caption{$4\pi r^2 g(r)$ as a function of $r$ at $\alpha=6$. From left to right curves (FVA after the formalism of 
Ref.~\protect\onlinecite{pee83} ) and points (PIMC) refers to 
 $\beta\hbar\omega_L=0.1, 1.0, 20$, respectively.}
 \label{fig:gdt}
\end{figure}

In Fig.~\ref{fig:gdt} we show the temperature dependence of $4\pi r^2 g(r)$
at intermediate coupling.

In Figs.~\ref{fig:gdr},\ref{fig:gdt} we see that, apart from a small
overestimate  of the polaron radius, FVA reproduces fairly well
the PIMC data. We can then safely use the 
FVA to discuss the spatial properties of the
polaron. Notice that the good agreement found between
PIMC and FVA is not  obvious since in principle
only the free energy is expected to be accurate for the latter. 

Contrary to the naive expectation the effect of 
temperature is to shrink the polaron (Fig.~\ref{fig:gdt}).
Our physical explanation is the
following: 
 At low temperatures phonons relax shrinking the spatial extent of the 
electron till the increase in electron kinetic energy balances the gain in
electron-phonon interaction energy. At high temperatures, however, a typical electron
has energy $\bar E=3/2\beta$ and momentum $\hbar \bar k = \sqrt{3 m/\beta}$.
 One can construct a wave packet of width $\Delta k$ in momentum space
using plane waves with higher and smaller energy without affecting the 
electron
internal energy. The biggest  $\Delta k$ which will not affect the electron
internal energy is of order of $\bar k$ itself. One can then achieve a 
localization 
of the electron of order $1/\bar k\sim  \hbar/ \sqrt{3 m/\beta}\sim \lambda_T$.
It follows that the phonons can relax at practically no cost  till the polaron 
radius  stabilizes at a value of this order. In fact
at high temperatures  the asymptotic value of the  polaron radius 
can be obtained from Eq.~(\ref{eq:ghiT}):
$\lim_{\beta\rightarrow 0}r_{m}=l\sqrt{\beta \hbar\omega_L}/2=0.2\lambda_T$.
This scaling has been found by Sethia {\it et al.}\cite{Set99}
for the mean square displacement of the electron in imaginary time.
Notice that the polaron radius becomes independent of the coupling.

The same behavior of $r_m$ has been obtained in 
Ref. \onlinecite{pee85} within FVA.  The authors of Ref.~\onlinecite{pee85} 
ascribe the high temperature behavior of the polaron radius to the 
increased fluctuations of the phonon field.  We conversely think that 
the thermal fluctuations of the electron  are responsible
of the high temperature behavior of the polaron radius and the 
phonon field acts only as a probe of the intrinsic  thermal
length of the electron as explained above.

\begin{figure}[htbp]
 \epsfxsize=8cm
   \epsfysize=5cm
 $$
 \epsfbox{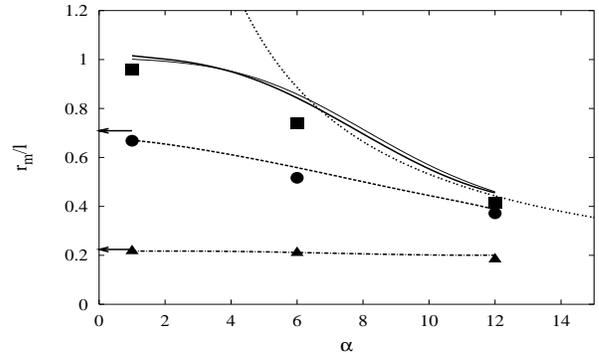}
 $$
 \caption{The polaron radius as a function of coupling for different
inverse temperatures. Results from FVA are shown as dash-dotted 
($\beta\hbar\omega_L=0.1$), 
dashed ($\beta\hbar\omega_L=1.0$),
thick-solid ($\beta\hbar\omega_L=20$)
and thin-solid lines ($\beta\hbar\omega_L=\infty$).
Dotted line is the Landau Pekar approximation. 
Results from PIMC calculations are shown as
triangles ($\beta\hbar\omega_L=0.1$),
circles ($\beta\hbar\omega_L=1.0$)
and squares ($\beta\hbar\omega_L=20$). 
The horizontal arrows indicate the value 
$0.2 \lambda_T/l$ for $\beta\hbar\omega_L=1$ (upper) and 
$\beta\hbar\omega_L=0.1$ (lower).
The arrows for $\beta\hbar\omega_L=20,\infty$ are out of scale.
}
 \label{fig:rda}
\end{figure}

To characterize the temperature and coupling dependence of the polaron
size we plot in 
Fig.~\ref{fig:rda}   the polaron radius as a function of 
coupling for different inverse temperatures.  
We see that when  the temperature is such that 
$0.2 \lambda_T>l$ (low temperatures, see the curves for $\beta\hbar\omega_L=20,\infty$),  
the polaron radius exhibits 
little temperature dependence at all couplings and simply interpolates 
between the weak coupling polaron 
radius $r_m=l$ and the Landau-Pekar polaron radius 
$r_m=3l\sqrt{\pi/2}/\alpha$.
  When $0.2 \lambda_T<l$ two different regimes occur. At small 
coupling the polaron radius tends to saturate at $0.2 \lambda_T$ 
(horizontal arrow) whereas
at high couplings one recovers the low temperature polaron radius
$r_m(\alpha,T=0)$. We can define a crossover line when these two lengths are
of equal magnitude so that the crossover temperature as a function of $\alpha$ is
given by the equation  $r_m(\alpha,T=0)= 0.2 \lambda_T(T^*)$. 
In the FVA approximation \cite{clp00} 
we find the crossover temperature in energy units to be 
$T^*(\alpha)=0.15 \hbar\omega_L v(\alpha,T=0)$ with $v$ the usual
Feynman variational parameter\cite{fey65}.  
For $\alpha \rightarrow 0$, $T^*$ goes to a constant of order of 
$0.5 \hbar\omega_L$
whereas for large $\alpha$ it increases quadratically with $\alpha$. 

In the low temperature
regime ($T<T^*$) the polaron radius becomes almost
temperature independent and is  determined by the coupling\cite{note}.
The high temperature regime $T>T^*$ is characterized by a polaron radius 
which is independent of the coupling and is determined by the 
temperature alone $(r_m=0.2 \lambda_T)$.


\section{Conclusions}

We have studied the charge induced in the lattice by 
an electron in  Fr\"ohlich polaron problem and the associated
polarization field.
We have derived relations which express the 
charge-induced density correlation function in terms of path integral involving
only the electronic degree of freedom which are suitable to be evaluated by
PIMC method.
A rigorous sum rule was derived that determines the total induced
charge and the long distance behavior of the polarization field.  
We give also the asymptotic limits of these quantities at short
distances and at high temperatures. 
We have compared results obtained using FVA 
through the lines of  Refs.~\onlinecite{pee83,pee82,pee85} and those obtained
by PIMC method. To the best of our knowledge this is the first 
PIMC computation of  real space correlation functions in  Fr\"ohlich model.
From the spatial dependence of the induced charge we obtained a polaron 
radius. 
The polaron radius  is determined by the coupling
at low temperatures and by the thermal wave length at high
temperatures with a crossover temperature that we evaluated in 
FVA. 

At high temperature a polaron with small
radius and small effective mass is achieved. 
These results are not in contradiction 
because the small radius at high temperatures
is a thermal effect of the electron and it is not related to the 
lattice response. The lattice acts only as a probe of the intrinsic
electron localization radius namely $\lambda_T$. 
Obviously this small radius 
polaron has nothing to do with the Holstein zero-temperature 
small polaron which induces almost local lattice displacements and 
moves coherently with a large
effective mass.

The PIMC polaron radius is always smaller than the FVA calculation
in the range of coupling and temperature studied. This effect 
is more pronounced at intermediate couplings. 
The overall temperature dependence agrees with the findings of 
Ref.~\onlinecite{pee83,pee82,pee85} however our physical interpretation is
different.

\acknowledgments
We acknowledge the third anonymous referee to bring us the attention on 
Ref.~\onlinecite{que95} (which suggested us the second proof of the
induced-density sum rule)
and Ref.~\onlinecite{Set99}. We acknowledge useful 
discussions and suggestions from S. Fratini and J. T. Titantah. J.L. 
thanks P. Calvani's group for hospitality during this work.  
We acknowledge partial support
from the MURST 1997 matching funds program
no.9702265437.


\begin{references}
\bibitem{note}
We remark that within  FVA and for weak coupling, one finds
a small increase of $r_m$ with temperature with a shallow 
maximum (close to  $\beta^{-1}\simeq0.15\hbar\omega_L$ for $\alpha=1$)
follow by a decrease at high temperatures according to 
the behaviour discussed in the text.
\end{references}
\end{document}